\documentclass{achemso}
\usepackage{cmap}
\usepackage[utf8]{inputenc}
\usepackage[T1]{fontenc} 
\usepackage{amssymb}          
\usepackage{amsfonts}          
\usepackage{textcomp}            
\usepackage{mathtools}
\usepackage{enumitem}           
\usepackage{hyperref}
\usepackage{amsmath}
\usepackage{upgreek}
\usepackage{color}
\usepackage{siunitx} 
\usepackage[german, USenglish]{babel}

\newcommand{\sub}{\mathcal{S}}
\newcommand{\enz}{\mathcal{E}}
\newcommand{\pro}{\mathcal{P}}
\newcommand{\kon}{k_\text{on}}
\newcommand{\koff}{k_\text{off}}
\newcommand{\kcat}{k_\text{cat}}
\newcommand{\km}{K_\text{M}}
\newcommand{\De}{D_\text{e}}
\newcommand{\Df}{D_\text{f}}
\newcommand{\Dc}{D_\text{c}}
\newcommand{\Ds}{D_\text{s}}
\newcommand{\Dp}{D_\text{p}}
\newcommand{\Rf}{R_\text{f}}
\newcommand{\vh}{\vec{\text{v}}_\text{h}}
\newcommand{\dv}{\delta \vec{\text{v}}}
\newcommand{\deltav}{\delta\vec{\text{v}}}
\newcommand{\vph}{v_\text{ph}}

\newcommand{\Dxd}{D_\text{xd}}
\newcommand{\phies}{\phi_\text{fs}}
\newcommand{\phics}{\phi_\text{cs}}
\newcommand{\Jxd}{J^\text{e}_\text{xd}}
\newcommand{\Jd}{J^\text{e}_\text{D}}

\newcommand{\sh}{s_\text{h}}
\newcommand{\eh}{e_\text{h}}
\newcommand{\ph}{p_\text{h}}
\newcommand{\sr}{s_\text{R}}
\newcommand{\lambdac}{\lambda_\text{c}}
\newcommand{\lambdaf}{\lambda_\text{f}}
\newcommand{\gammas}{\gamma_\text{s}}
\newcommand{\gammap}{\gamma_\text{p}}
\newcommand{\lambdacf}{\lambda_\text{c/f}}
\newcommand{\Cc}{C_\text{c}}
\newcommand{\Cf}{C_\text{f}}
\newcommand{\Ccf}{C_\text{c/f}}
\newcommand{\tensor}[1]{\textbf{#1}}

\author{Giovanni Giunta}
\author{Hamid Seyed-Allaei}
\author{Ulrich Gerland}
\email{gerland@tum.de}
\affiliation[TUM]
{Physics Department, Technical University of Munich, Garching, Germany}

\title{Cross-diffusion induced patterns for a single-step enzymatic reaction}


\begin{document}
	

\begin{abstract}
Several different enzymes display an apparent diffusion coefficient that increases with the concentration of their substrate. Moreover, their motion becomes directed in substrate gradients. Currently, there are several competing models for these transport dynamics. Here, we analyze whether the enzymatic reactions can generate a significant feedback from enzyme transport onto the substrate profile. We find that this feedback can generate spatial patterns in the enzyme distribution, with just a single-step catalytic reaction. However, patterns are formed only for a subclass of transport models. For such models, nonspecific repulsive interactions between the enzyme and the substrate cause the enzyme to accumulate in regions of low substrate concentration. Reactions then amplify local substrate fluctuations, causing enzymes to further accumulate where substrate is low. Experimental analysis of this pattern formation process could discriminate between different transport models. 
\end{abstract}


\section{Introduction}
\label{sec:intro}

Experiments performed during the last decade found that at least eight different enzymes display a higher diffusion coefficient when the concentration of the corresponding substrate in solution is increased \cite{borsch_conformational_1998, muddana_substrate_2010, riedel_heat_2015, illien_exothermicity_2017, jee_enzyme_2018, zhao_substrate-driven_2018, yu_molecular_2009}. These increases are in the range of $24\text{-}80\%$ relative to the diffusion coefficients without substrate.
Most of these experiments relied on fluorescence correlation spectroscopy (FCS) measurements. Although artifacts introduced by this technique have been pointed out \cite{gunther_diffusion_2018, zhang_aldolase_2018}, recent findings using other techniques \cite{jee_enhanced_2019, xu_direct_2019} validated the phenomenon, which is often referred to as ``enhanced diffusion''. 
The underlying mechanism is still under debate \cite{zhang_enhanced_2019}. Some experiments suggest that catalysis plays a key role \cite{riedel_heat_2015, jee_enzyme_2018, jee_catalytic_2018, bonnin_mobility_2019}, while others indicate that enhanced diffusion persists when the substrate is replaced by an inhibitor \cite{borsch_conformational_1998, illien_exothermicity_2017, mohajerani_theory_2018}. According to the latest experiments performed with the enzyme urease \cite{jee_catalytic_2018,xu_direct_2019}, it appears that both the binding and the catalysis step of the reaction scheme contribute to enhanced diffusion.  

A related question is how enzymes behave in the presence of substrate gradients. The answer to this question appears to be complex. Some experiments suggest that enzymes drift downstream gradients of substrates, performing ``antichemotaxis'' \cite{jee_enzyme_2018,jee_catalytic_2018}. Others suggest that enzymes move upstream gradients of substrates, performing ``chemotaxis'' \cite{zhao_substrate-driven_2018,mohajerani_theory_2018,yu_molecular_2009}. 
Antichemotaxis can be explained based on just the enhanced diffusion \cite{jee_enzyme_2018,weistuch_spatiotemporal_2018} (enzymes accumulate in regions with low substrate concentration where they have a lower diffusion coefficient), chemotaxis cannot be generated by enhanced diffusion alone. 
However, cross-diffusion is a possible cause for enzyme chemotaxis \cite{zhao_substrate-driven_2018}. Cross-diffusion describes the response of the enzyme to forces generated by gradients of substrate. Mathematically, it corresponds to an off-diagonal element in the diffusion matrix describing the combined motion of the enzyme and the substrate. It has been suggested that cross-diffusion can be due to specific interactions (ligand binding) between the enzyme and the substrate \cite{mohajerani_theory_2018, schurr_theory_2013, zhao_substrate-driven_2018} or due to nonspecific interactions (e.g. steric, van der Waals) \cite{agudo-canalejo_phoresis_2018}.
Specific interactions only lead to chemotaxis, while nonspecific interactions can cause the enzymes to move both up- or downstream the substrate gradient, depending on whether the interactions are attractive or repulsive, respectively. The model including nonspecific interactions~\cite{agudo-canalejo_phoresis_2018} can be considered as a mathematical generalization of other existing models \cite{jee_enzyme_2018, schurr_theory_2013, zhao_substrate-driven_2018}.

While the existing models and experiments study how enzymes move in pre-imposed substrate gradients, they do not consider the feedback from the enzymatic reaction onto the substrate distribution. Here, we analyze the effects of this feedback starting from the most general transport model \cite{agudo-canalejo_phoresis_2018}. We show that spatial patterns can emerge in initially homogeneous systems if nonspecific interactions contribute to the accumulation of the enzyme in regions where concentration of substrate is low. 
Enzymes accumulating in these regions further deplete the substrate, causing the substrate gradient to become steeper, hence further increasing the accumulation of the enzyme. 
We obtain a set of conditions for the parameter range in which patterns form.
We see that patterns can emerge only for repulsive nonspecific interactions, hence only for the model proposed in Ref.~\cite{agudo-canalejo_phoresis_2018}, but not for the models proposed in Ref.~\cite{zhao_substrate-driven_2018,mohajerani_theory_2018} and Ref.~\cite{jee_enzyme_2018}, suggesting that the analysis of pattern formation experiments can be used to discriminate between the different proposed models.
Our findings imply that patterns can arise for a single-step enzymatic reaction even in the absence of autocatalytic activity or allosteric regulation. This is surprising given that the formation of conventional Turing patterns \cite{turing_chemical_1990} with such simple reaction schemes requires at least a reaction network of three states, where forward and backward reactions are catalyzed by two different enzymes \cite{sugai_design_2017}.

\section{Results}
\label{sec:results}

\begin{figure}[!t]
	\centering
	\includegraphics{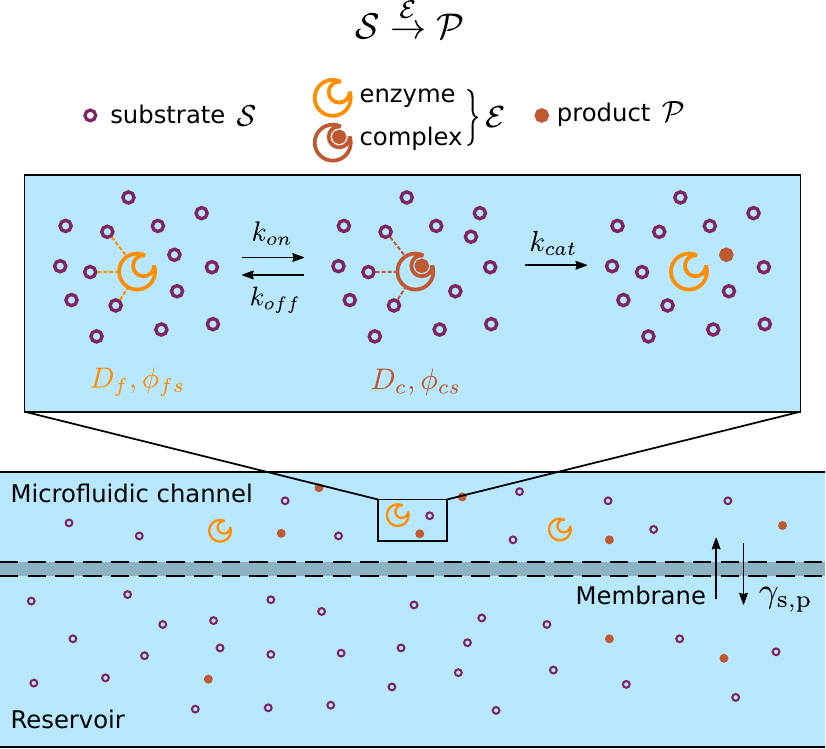}
	\caption{System considered in this work. A single step enzymatic reaction takes place in a narrow reaction chamber which is coupled to a reservoir through a permeable membrane. The membrane allows for the exchange of only substrates and products with rate $\gammas$, $\gammap$ respectively, but no exchange for the enzymes. Inset: Reaction scheme. The substrate $\sub$ is converted into a product $\pro$ by an enzyme $\enz$. The reaction follows a Michaelis-Menten scheme where substrate can bind to the free enzyme with rate $\kon$, forming a complex, and unbind with rate $\koff$. The catalytic step of the reaction has a rate $\kcat$. We assume that the enzyme has a diffusion constant $\Df$, $\Dc$ depending on whether the enzyme is free or in its complexed form. Nonspecific pairwise interactions $\phies$, $\phics$ can depend on the enzyme form.
	}\label{fig:model}
\end{figure}

Our model system is depicted in Fig.~\ref{fig:model}. We consider a single-step enzymatic reaction in a narrow reaction chamber connected via a permeable membrane to a large substrate reservoir. We assume only substrate and product molecules can diffuse through the permeable membrane (exchange rates $\gammas$, $\gammap$, respectively), while enzymes are confined to the reaction chamber (since enzymes are typically larger than their substrates and products). The reservoir has a fixed concentration of substrate $\sr$ and no products.
For the reaction occurring in the bulk of this effectively 1D system we assume a Michaelis-Menten scheme (Fig.~\ref{fig:model}, inset). The substrate $\sub$ binds to the enzyme $\enz$ with rate $\kon$, forming a complex, and it can unbind with rate $\koff$. The catalytic step of the reaction has rate constant $\kcat$ and catalysis is irreversible. This leads to a turnover rate per enzyme $\kcat F(s):=\kcat s/(\km+s)$, where $s$ is the substrate concentration and $\km=(\koff+\kcat)/\kon$. 
Following Ref.~\cite{agudo-canalejo_phoresis_2018} we assume that the diffusion coefficient of the enzyme depends on whether the enzyme is free, $\Df$, or it is in its complexed form, $\Dc$. This change in diffusion coefficient can be due to changes in either the hydrodynamic radius or the conformational fluctuations of the enzyme upon substrate binding \cite{illien_exothermicity_2017, illien_diffusion_2017, kondrat_brownian_2019}. Typically, enzymes become more tightly folded upon substrate binding, i.e. $\Dc >\Df$, consistent with the experimentally observed trend.
We also assume that short-range nonspecific interactions can either cause the enzyme to move towards or away from substrate, effectively generating a phoretic drift velocity that can also be interpreted as cross-diffusion \cite{agudo-canalejo_phoresis_2018}. We denote these interactions via pairwise potentials $\phies$, $\phics$ depending on whether the enzyme is free or in the complexed form, respectively, as depicted in Fig.~\ref{fig:model}(inset).

Under the assumptions that (i) the enzyme is very dilute and (ii) the system is locally in chemical equilibrium (i.e. the timescales of diffusion and cross-diffusion are slower than the chemical reactions), one can derive an effective transport equation for the enzymes~\cite{agudo-canalejo_phoresis_2018},
\begin{align}\label{eq:e_motion}
	\partial_t e(x,t) & =  \partial_x^2 [\De(s) e] +\partial_x [\Dxd(s,e) \partial_x s] 
\end{align}
within the quasi-1D reaction chamber (Fig.~\ref{fig:model}) oriented along the $x$ axis. Here, $e(x,t)$ denotes the local enzyme concentration (regardless of free or complexed) and $s(x,t)$ the substrate concentration. 
The effective diffusion coefficient of the enzyme, $\De(s)=\Df+(\Dc-\Df)F(s)$, is a function of substrate concentration, interpolating between the diffusion coefficient of the free enzyme and the complex, with $F(s)$ as defined above. For $s \ll \km$, $\De(s)\sim \Df$, whereas $\De(s)\sim \Dc$ for $s \gg \km$. Note that the disassembly of enzyme oligomers into monomers can also contribute to enhanced diffusion~\cite{shah2013f1, zhang_aldolase_2018, jee_enhanced_2019}. Eq.\eqref{eq:e_motion} would then describe the motion of the enzyme irrespective of its oligomeric state.
The cross-diffusion term $\Dxd(s,e)$ of Eq.~\eqref{eq:e_motion} describes how enzymes respond to gradients of substrate due to the short-range nonspecific and hydrodynamic interactions, 
\begin{equation}\label{eq:vph}
	\Dxd(s,e) = -\left[ \Cf + (\Cc -\Cf) F(s) \right]e \,. 
\end{equation}
Here, $\Ccf= N_A k_B T \lambdacf^2/\eta $, where $\eta$ is the viscosity of the fluid, $k_B$ the Boltzmann constant, $N_A$ the Avogadro number, $T$ the temperature and $\lambdacf$ the Derjaguin length \cite{derjaguin1947kinetic,agudo-canalejo_phoresis_2018}. 
The Derjaguin length is a parameter capturing the effective short range interaction between the complex/free enzyme and the substrate. It is typically a few angstroms \cite{ebbens_size_2012,anderson1989colloid}, which is smaller than the Debye length (screening length) in typical buffer conditions ($\approx 1nm$)~\cite{zhang_enhanced_2019}. It is expressed via the integral $\lambdacf^2=\int_0^{\infty} dhh(e^{-\phi_{cs/fs}(h)/(k_B T)}-1)$. $\lambdacf^2$ is positive (negative) when the interaction is attractive (repulsive) \cite{agudo-canalejo_phoresis_2018}. The derivation of $\lambdacf^2$ is similar to that of the second Virial coefficient for a real gas~\cite{mcquarrie_statistical_1975}, but it also  includes hydrodynamic corrections and is computed by assuming that the size of the enzyme is much larger than the interaction length.
The sign of $\lambdacf^2$ determines the sign of $\Ccf$ and ultimately the sign of $\Dxd(s,e)$. For attractive interactions $\Dxd(s,e)<0$, i.e. the enzyme drifts towards higher concentrations of substrate (chemotaxis). The enzyme performs antichemotaxis for repulsive interactions. Note that the effect of nonspecific interactions can also be written as a phoretic drift~\cite{agudo-canalejo_phoresis_2018} by swapping the $\partial_x s$ in Eq.~\eqref{eq:e_motion} with $e$ in the definition~\eqref{eq:vph}, with a drift velocity directly proportional to the substrate gradient $\vph(s,\partial_x s)=\left[ \Cf + (\Cc -\Cf) F(s) \right]\partial_x s$.

In the regime where the enzyme is dilute and Eq.~\eqref{eq:e_motion} is valid, the reaction chamber of Fig.~\ref{fig:model} is then described by the coupled reaction-transport equations  
\begin{align}\label{eq:RDE_system}
	\begin{cases}
	\partial_t e(x,t) = \partial_x^2 [\De(s) e] +\partial_x [\Dxd(s,e) \partial_x s] \\
	\partial_t s(x,t) = \Ds \partial_x^2 s -\kcat e F(s) -\gammas [s-\sr] \\
	\partial_t p(x,t) = \Dp \partial_x^2 p +\kcat e F(s)-\gammap p\,,
	\end{cases}
\end{align}
where $p(x,t)$ denotes the product concentration. While the product dynamics does not influence the substrate and enzyme equations, we include it as spatial read-out of the reaction. Having the reaction chamber coupled to a reservoir avoids product accumulation and substrate depletion, generating a nonzero homogeneous steady-state, with concentrations $\eh$, $\sh$, and $\ph$ for enzyme, substrate, and product, respectively. For simplicity, we express $\eh$ and $\ph$ as functions of $\sh$,  \begin{align}
\eh = \gammas \frac{\sr - \sh}{\kcat F(\sh)}, \label{eq:eh}
\\
\ph = \frac{\kcat \eh F(\sh)}{\gammap}\,. \label{eq:ph}
\end{align}
Since $F(\sh)$ is a monotonic increasing function of $\sh$, it is possible to write $\sh$ and $\ph$ in terms of $\eh$, which can be directly tuned in experiments via the total enzyme concentration  (see \hyperref[SI_sec:LIA]{SI} for the full expressions).

\begin{figure}[t]
	\centering
	\includegraphics{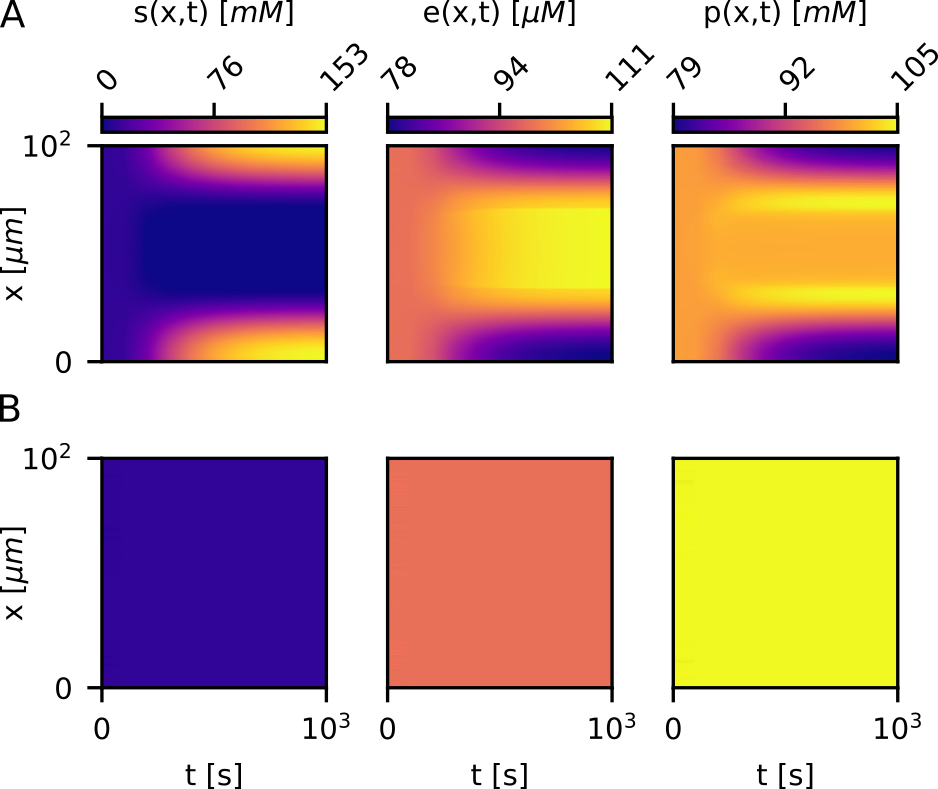}
	\caption{The concentration profiles for $s(x,t)$, $e(x,t)$, $p(x,t)$ are plotted as a function of time and spatial coordinate for two different parameter sets (see Table S1 for the full parameter list). (A) ($\sh=10^4 \mu M$,$\lambda^2=-1 \si{\angstrom}^2$): spatial patterns arise from initial homogeneous concentrations with white gaussian noise. We can see how the enzyme accumulates in regions of low substrate concentration, as well as the product; the profiles reach the steady-state after about $\sim 10^3$ seconds. (B) ($\sh=10^4 \mu M$,$\lambda^2=1 \si{\angstrom}^2$): the initial perturbation decays and the steady state profiles are homogeneous.}
	\label{fig:patterns}
\end{figure}

The homogeneous solution as given by Eqs.~\eqref{eq:eh},\eqref{eq:ph} is stable for any positive values of the parameters for a well-mixed system, i.e. a system with no diffusion and no cross-diffusion (see \hyperref[SI_sec:LIA]{SI}). Fig.~\ref{fig:patterns} shows the results of two simulations of the full system, Eq.~\eqref{eq:RDE_system}, with periodic boundary conditions and parameters as given in Table S1. In Fig.~\ref{fig:patterns}A we see that the homogeneous solution is unstable and patterns form, for a value of $\lambdaf^2=\lambdac^2=-1\si{\angstrom}^2$. In Fig.~\ref{fig:patterns}B we see that for $\lambdaf^2=\lambdac^2=1\si{\angstrom}^2$ the homogeneous solution is instead stable. Hence, depending on the parameters, the system will spontaneously form patterns .

To characterize the instability of the homogeneous steady state solution, $\vh=(\eh,\sh,\ph)$, we linearize Eqs.~\eqref{eq:RDE_system}, $\vec{\text{v}}\rightarrow\vh+\deltav$, and make the exponential ansatz $\deltav=\vec{\text{v}}_0 \, e^{\sigma t}e^{i q x}$, with $q$ the spatial frequency of the linear perturbation and $\sigma$ the perturbation growth rate. If $\sigma>0$ the perturbation grows with time and patterns form.
A positive $\sigma$ can be found provided that (see \hyperref[SI_sec:LIA]{SI})
\begin{eqnarray}
& \Df +\sh \left( \Cf+ \frac{\sh}{\km} \Cc \right)<0\,, \label{eq:inst_cond_Df}
\\
& 0 < q < \sqrt{-\frac{\beta [\Df \km +\sh(\Cf \km+\Cc \sh)]}{\Ds \sh(\Df\km +\Dc \sh)}}\,,  \label{eq:inst_cond_q}
\end{eqnarray}
with $\beta=\gammas \sr$. This result holds in the strong depletion regime ($\sh \ll \sr$), where the effect of the reaction on the substrate dominates over the outflow to the reservoir. In this regime, the expressions are simpler and it is easier to pinpoint the driving mechanism behind the instability observed in Fig.~\ref{fig:patterns}. We refer the reader to the \hyperref[SI_sec:LIA]{SI} for the full analysis. 
The argument of the square root in the inequality~\eqref{eq:inst_cond_q} is positive if relation~\eqref{eq:inst_cond_Df} is fulfilled. In the \hyperref[SI_sec:LIA]{SI} we show that the instability is a Type II instability \cite{cross_pattern_2009}, meaning that $\sigma=0$ at $q=0$. This is natural as the total amount of enzymes in our system is conserved and homogeneously increasing or decreasing perturbations, i.e. perturbations at $q=0$, would correspond to changes in the total enzyme amount.
By looking at the inequalities~\eqref{eq:inst_cond_Df}, \eqref{eq:inst_cond_q}, we can see that repulsive interactions, i.e. $\Ccf<0$, are needed to have instabilities. Diffusion tends to homogenize the concentration profiles and contributes with a positive term to the left hand side of inequality~\eqref{eq:inst_cond_Df}. 
In the case where the enzyme-substrate interaction is attractive, i.e. $\Cf>0$, also the second term on the left hand side of inequality~\eqref{eq:inst_cond_Df} is positive. Hence the interaction between the enzyme and the substrate would need to change sign upon substrate binding to have an instability ($\Cc<0$). 
There can be cases in which the change in nonspecific interactions is less abrupt and both $\Cf<0$ and $\Cc<0$.
Even in the case for which $\Cc = \Cf = C$, i.e. there is no change in interaction upon substrate binding, it is possible to have an unstable homogeneous solution for 
\begin{equation}\label{eq:cond_constant_lambda}
\lambda^2 < -\frac{1}{6 \pi \Rf N_A}\frac{\km}{\sh(\sh+\km)}\,,
\end{equation}
where we rewrote relation~\eqref{eq:inst_cond_Df} with the use of the Stokes-Einstein relation, $k_B T=6 \pi \eta \Df \Rf$, and the definition of $C$.
By considering biologically relevant ranges, such as $\Rf\sim 1\text{-}10 nm$~\cite{arrio-dupont_translational_2000,erickson_size_2009}, $\km\sim 10^{-2}\text{-}10^3 \mu M$~\cite{bar-even_moderately_2011}, we find that the critical value to be in the unstable regime $|\lambda^*|$ is smaller than the Debye length $|\lambda^*|\approx 10^{-2}\text{-}10\si{\angstrom}$, for $\sh=10 m M$.

\begin{figure}[t]
	\centering
	\includegraphics{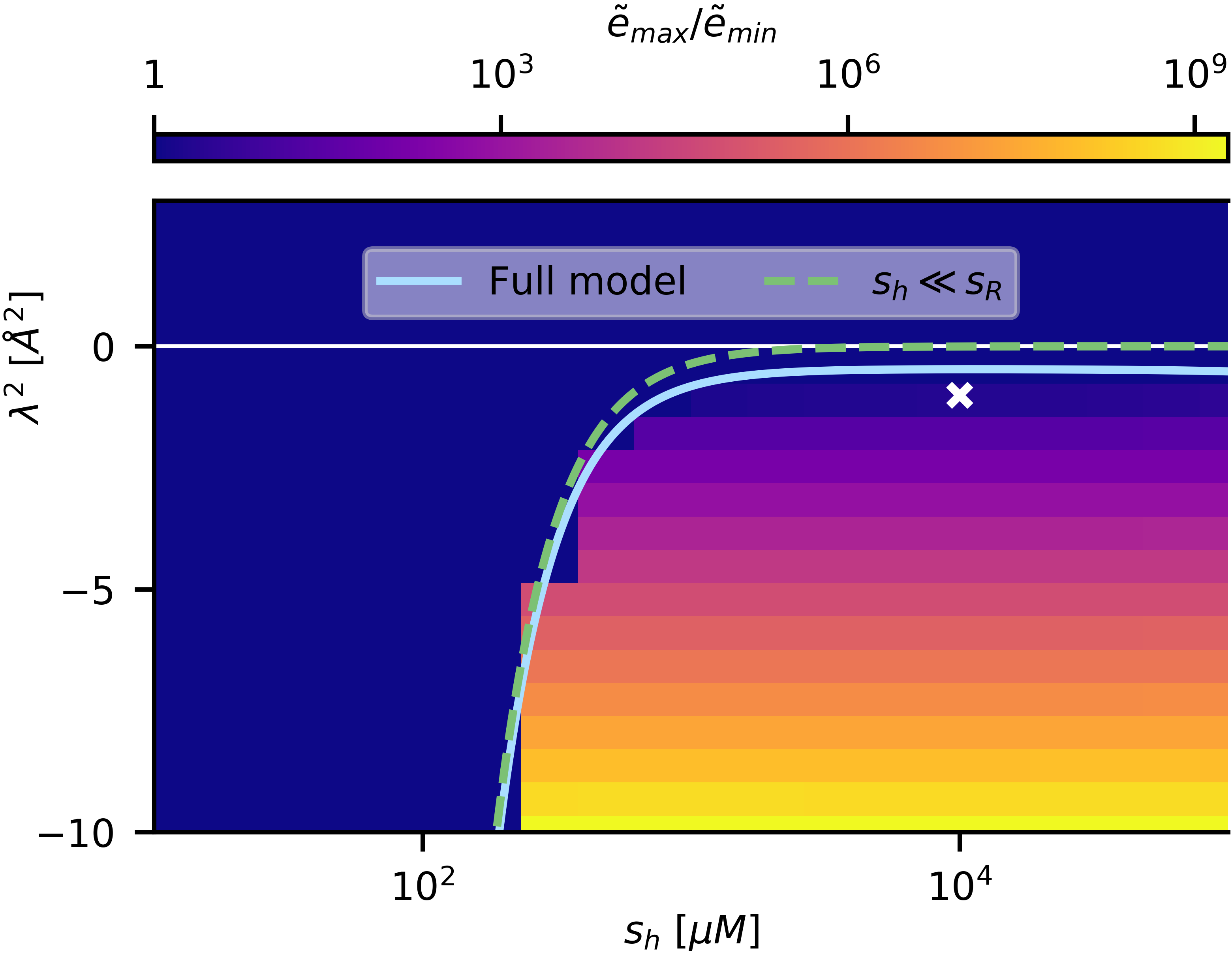}
	\caption{Phase diagram where we plot the instability curve of the model~\eqref{eq:RDE_system} (light blue line) and the approximated curve as given by relation~\eqref{eq:cond_constant_lambda} (green dashed line). As a proxy for patterns, we plot the ratio of the maximum of the steady state profile of the enzyme concentration over the minimum. Below the curve, the system \eqref{eq:RDE_system} is unstable and patterns arise, the ratio $\tilde{e}_{max}/\tilde{e}_{min}$ can be as high as $10^9$. Above the curve, the homogeneous solution is stable and the ratio $\tilde{e}_{max}/\tilde{e}_{min} =1$. The white cross corresponds to the simulation shown in Fig.~\ref{fig:patterns}A. For the full parameter list we refer the reader to the \hyperref[SI_sec:params]{SI}.}
	\label{fig:phase_diagram}
\end{figure}

It is interesting to note how relations \eqref{eq:inst_cond_Df}-\eqref{eq:cond_constant_lambda} depend on the substrate concentration. One could ask, given certain nonspecific interactions between the enzyme and the substrate, at which substrate concentration $\sh^*$ should we begin to observe instabilities? 
From relations \eqref{eq:inst_cond_Df}- \eqref{eq:cond_constant_lambda}, we find that $\sh^*=\sqrt{\Df\km/|C|}=\sqrt{\km/(6\pi\Rf N_A |\lambda^{2*}|)}$, with $\lambda^{2*}<0$, i.e. for repulsive interactions, and where we considered the saturated regime $\sh \gg \km$, i.e. $F(s)\approx 1$.
In Fig.~\ref{fig:phase_diagram} we plot the phase diagram of the system of Eq.~\eqref{eq:RDE_system} where on the abscissa we have $\sh$ and on the ordinate we have $\lambda^2$. The green dashed line corresponds to the instability curve~\eqref{eq:cond_constant_lambda} in the strong depletion regime ($\sh \ll \sr$) and the light blue line represents the instability curve derived for any $\sh$ (see \hyperref[SI_sec:LIA]{SI}). As a proxy for the determination of patterns we plot the ratio of the maximum over the minimum of the enzyme profile at steady-state. Above the instability lines the homogeneous solution is stable. Below the line the system is unstable and patterns similar to the one shown in Fig.~\ref{fig:patterns}A arise.  

\begin{figure}[t]
	\centering
	\includegraphics{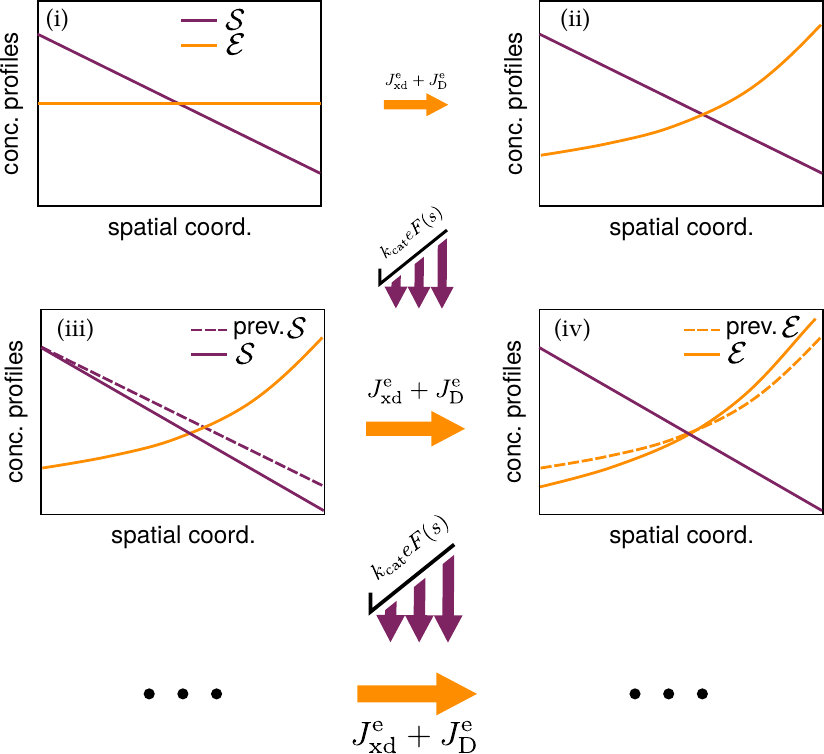}
	\caption{Positive feedback mechanism behind the pattern forming process. (i) We start with a substrate gradient and a homogeneous enzyme profile; (ii) the enhanced diffusive current $\Jd$ and the cross-diffusive one $\Jxd$ cause the enzyme to accumulate where substrate is low; (iii) reaction causes the substrate gradient to get steeper; (iv) this leads to a further increase of $\Jd$ and $\Jxd$ causing further accumulation of the enzyme. This process repeats itself determining the patterns. }
	\label{fig:intuition}
\end{figure}

What is the physical mechanism underlying the instability given by the inequalities \eqref{eq:inst_cond_Df},\eqref{eq:inst_cond_q}? We illustrate the feedback mechanism generating the pattern in Fig.~\ref{fig:intuition}.
One can show that if relation \eqref{eq:inst_cond_Df} is fulfilled, the cross-diffusion $\Dxd(s,e)>0$. The enzymatic current induced by the cross-diffusion is given by $\Jxd=-\Dxd(s,e)\partial_x s$, which for a negative slope of substrate concentration generates a positive current for the enzyme, i.e. the enzyme moves away from a high substrate concentration. The current generated by the enhanced diffusion $\Jd=-\partial_x[\De(s)e]$ also consists of a motion of the enzyme away from high substrate concentrations.
In Fig.~\ref{fig:patterns}A we can see how in a regime where patterns form, more product is generated in locations where the substrate concentration is low and the enzyme concentration is high. Repulsive nonspecific interactions and enhanced diffusion cause the enzyme to accumulate in such regions  (Fig.~\ref{fig:intuition} from (i) to (ii)). This accumulation then generates a higher reaction flux in these regions. Having a stronger reaction flux where substrate is already low, as compared with regions where substrate is abundant, causes substrate gradients to become steeper (Fig.\ref{fig:intuition} from (ii) to (iii)). A steeper substrate gradient, in turn, causes both $\Jxd$ and $\Jd$ to increase, hence generating a further accumulation of the enzyme in substrate depleted regions (Fig.~\ref{fig:intuition} from (iii) to (iv)). This positive feedback between reaction and enzyme accumulation, leads to the formation of patterns. The feedback cycle halts when the substrate concentration is too low and the reaction is balanced by the influx of substrate from the reservoir. Then the substrate gradient stops getting steeper and the system approaches a steady-state.

What happens without enhanced diffusion, with only nonspecific interactions? The diffusion function $\De(s)$ in Eq.~\eqref{eq:RDE_system} becomes a constant, i.e. $\De=\Dc$, and the interval~\eqref{eq:inst_cond_q} for the unstable wave vector $q$ is slightly affected. The inequality~\eqref{eq:inst_cond_Df} characterizing the instability is unaffected.
This result suggests that enzyme patterns for a single step reaction form only if driven by repulsive nonspecific interactions.  
Why can't patterns be generated simply by enhanced diffusion? If $\Dxd(s,e)=0$ in Eq.~\eqref{eq:RDE_system}, it is possible to observe patterns only if (see \hyperref[SI_sec:No_short_range]{SI}) 
\begin{equation}\label{eq:inst_cond_no_phoresis}
\frac{1}{\De(\sh)}\left. \frac{\partial \De(s)}{\partial s}\right|_{\sh}-\frac{1}{F(\sh)}\left. \frac{\partial F(s)}{\partial s}\right|_{\sh}>\frac{q^2 \Ds}{\beta} \,.
\end{equation}
Both $\De(s)$ and $F(s)$ have a Michaelis-Menten dependence on $s$ for all the models proposed so far for the enzyme motion. They differ only in the prefactors and a non-zero offset for $\De(s)$. Inequality \eqref{eq:inst_cond_no_phoresis} is never fulfilled for such $\De(s)$ and $F(s)$ and consequently any initial perturbation of the concentrations is smoothed out by diffusion.

However, the inequality \eqref{eq:inst_cond_no_phoresis} will apply to any model with a spatially dependent diffusion coefficient that is coupled to another diffusing and reacting species. Systems of this type are used to study bacterial motion \cite{liu_sequential_2011}. 
Interestingly synthetic bacterial populations show stripe patterns as they grow on semi solid agar plates \cite{liu_sequential_2011}. For this system $\De(s)$ and $F(s)$ have different functional forms.
The inequality \eqref{eq:inst_cond_no_phoresis} specifies the minimal ingredients for pattern formation for such systems.
It implies that patterns form whenever $\De(s)$ is more sensitive than $F(s)$ to perturbations in the substrate concentration. Having a more sensitive $\De(s)$ than $F(s)$ causes a more sensitive response in the enzyme motion than the depletion due to the reaction.
Consider a local increase in substrate concentration and that both $\De(s)$ and $F(s)$ are monotonically increasing functions of $s$. Having a more sensitive $\De(s)$ than $F(s)$ implies a higher increase in the current $\Jd$ due to enhanced diffusivity away from the substrate, as compared to the depletion of substrate due to the reaction. Molecules $\enz$ then migrate to regions with low substrate and if they do so with a high enough rate they can cause substrate gradients to get steeper, as in step (iii) of Fig~\ref{fig:intuition}. 
However, for the enzyme model \eqref{eq:RDE_system}, $\De(s)$ and $F(s)$ alone cannot generate instabilities. The accumulation of the enzyme in low substrate region at a high enough rate can be guaranteed only via the cross-diffusive term $\Dxd(s,e)$.

\section{Discussion}
\label{sec:discussion}

We have seen that patterns can form for a single-step catalytic reaction if cross-diffusive effects are present. Patterns form given sufficiently strong repulsive nonspecific interactions between the enzyme and the substrate, as indicated by the inequality~\eqref{eq:inst_cond_Df}. Repulsive interactions cause the enzyme to move away from regions of high substrate concentrations and to accumulate in regions of low substrate, performing antichemotaxis. The accumulated enzymes then deplete the substrate, steepening substrate gradients. Steeper gradients further drive the accumulation of enzymes as illustrated in Fig.~\ref{fig:intuition}. This positive feedback cycle between enzyme accumulation and reaction is what generates the patterns.

The enzyme accumulation is driven by the antichemotaxis due to nonspecific repulsive interactions. The enzyme chemotaxis considered in some of the models \cite{mohajerani_theory_2018,schurr_theory_2013,zhao_substrate-driven_2018} has a stabilizing effect. Enzymes accumulate in regions of high substrate concentrations. Then reactions flatten substrate gradients, breaking the feedback that leads to patterns. Hence for such systems patterns cannot form.
In the absense of short-range repulsive interactions, antichemotaxis still exists due to enhanced diffusivity \cite{jee_catalytic_2018,jee_enzyme_2018}. Nevertheless, we have seen that enhanced diffusivity alone cannot generate patterns for a simple enzymatic reaction.
Hence, among all models proposed so far for the enzyme motion, only the model given by Eq.~\eqref{eq:e_motion}, first proposed in Ref.~\cite{agudo-canalejo_phoresis_2018}, can lead to pattern formation. More generally, we believe that studying how the enzyme and the substrate affect each other on a macroscopic scale, can shed light on the microscopic mechanisms of enhanced diffusion and enzyme chemotaxis/antichemotaxis.  
The enzyme transport equation~\eqref{eq:e_motion} is qualitatively consistent with all experiments, because it can generate both chemotaxis and antichemotaxis.
However, Eq.~\eqref{eq:e_motion} needs to be further validated. Whether the enhanced diffusion can be expressed as $\De(s)=\Df+(\Dc-\Df)F(s)$ with $\Dc >\Df$, as first suggested in Ref.\cite{illien_exothermicity_2017,illien_diffusion_2017}, needs to be experimentally verified. 

The patterns observed in Fig.~\ref{fig:patterns} are not generated via the common short-range activation and long-range inhibition mechanism \cite{koch_biological_1994}, as neither the enzyme nor the substrate have autocatalytic activity. It is also not a motility induced phase separation (MIPS) mechanism \cite{cates_motility-induced_2015}, which relies on the slowing down of active particles in regions of high particle concentrations. Here a positive feedback mechanism between particle accumulation and reaction leads to the pattern formation. Note that for the system of equations~\eqref{eq:RDE_system}, patterns can form even if $\De\sim\Ds$, because inequality~\eqref{eq:inst_cond_Df} does not depend on $\Ds$, whereas for classical Turing patterns large differences in the diffusion coefficients of the different species are required \cite{turing_chemical_1990}. 
Moreover, it is surprising to see that patterns can form for a single-step enzymatic reaction with no autocatalytic activity nor allosteric regulation.
In fact, for a system where species have a constant diffusion coefficient and enzymatic reactions follow a simple Michaelis-Menten scheme, patterns form for a minimal network of three states, where forward and backward reactions are catalyzed by two different enzymes respectively \cite{sugai_design_2017}. In our system patterns form for a single step catalytic reaction because of cross-diffusion.

Our findings are consistent with recent studies analyzing the effects of cross-diffusion in pattern formation \cite{vanag_cross-diffusion_2009}. Moreover it has been shown that phase separation, formation of static or self-propelled aggregates can be observed in mixtures of cross-diffusive species interacting via a fast diffusing chemical that can be produced or consumed \cite{agudo-canalejo_active_2019}. 
The scenario considered here corresponds to the case of a single cross-diffusive species, i.e. the enzyme, that is able to consume the fast diffusing chemical, i.e. the substrate. Here we considered the full nonlinear forms of enhanced diffusion and cross-diffusion for the enzyme motion, whereas these other studies considered constant diffusion and constant cross-diffusion \cite{vanag_cross-diffusion_2009,agudo-canalejo_active_2019}.
The nonlinear model permitted us to address the question why enhanced diffusion alone is not able to generate patterns for a simple enzymatic reaction. We found that enhanced diffusion would need to be more sensitive to perturbations in substrate concentrations than the reaction, see Eq.~\eqref{eq:inst_cond_no_phoresis}. Although this is not the case for enzymes, we believe that relation~\eqref{eq:inst_cond_no_phoresis} can characterize the pattern formation of species presenting different enhanced diffusion functions, such as bacteria.


\begin{acknowledgement}
This work was funded by the Deutsche Forschungsgemeinschaft (DFG) within the framework of the Transregio 174 ``Spatiotemporal dynamics of bacterial cells". 
G.G. was supported by a DFG Fellowship through the Graduate School of Quantitative Biosciences Munich (QBM).
	
\end{acknowledgement}

\bibliography{ref.bib}

\newpage
\begin{center}
\Huge{\bf{Supplementary Information}}
\end{center}

\section{Linear Instability Analysis}
\label{SI_sec:LIA}
In this section, we perform the linear instability analysis of the PDE system~(3) of the main text to derive the condition for which patterns form. As a reminder, the starting model for our analysis (Eq~(3) of the main text) is:
\begin{equation}\label{eq_SI:RDE_system}
\begin{aligned}
	\begin{cases}
	&\partial_t e(x,t) = \partial_x^2 [\De(s) e(x,t)] + \partial_x [\Dxd(s,e)\partial_x s(x,t)] \\
	&\partial_t s(x,t) = \Ds \partial_x^2 s(x,t) -\kcat e(x,t) F(s) - \gammas (s - \sr) \\
	&\partial_t p(x,t) = \Dp \partial_x^2 p(x,t) + \kcat e(x,t) F(s)- \gammap p \,
	\end{cases}
\end{aligned}
\end{equation}
where $\Dxd(s,e) = -[ \Cf + (\Cc - \Cf) F(s) ] e(x,t)$, with $\Ccf= N_A k_B T \lambdacf^2/\eta $, $\eta$ is the viscosity of the fluid, $k_B$ the Boltzmann constant, $N_A$ the Avogadro's number, $T$ the temperature and $\lambdacf$ is the Derjaguin length \cite{derjaguin1947kinetic,agudo-canalejo_phoresis_2018}.

\subsection{Homogeneous steady state solution}

\begin{figure}[!htb]
	\centering
	\includegraphics{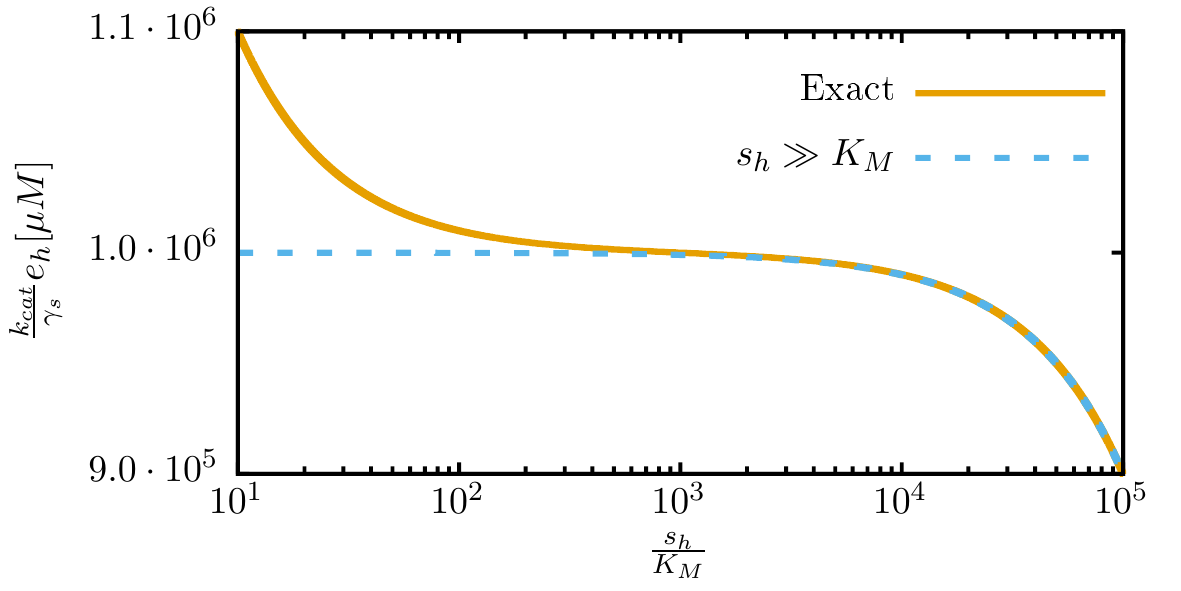}
	\caption{The relation between $\eh$ and $\sh$. Exact curve is for Eqs.\eqref{eq_SI:eh} or \eqref{eq_SI:sh} and the curve for $\sh \gg \km$ is for Eq.\eqref{eq_SI:eh-approx}. Parameters are $\sr = 10^6 \mu M$ and $\km = 1 \mu M$.}
	\label{fig:e-s-relation}
\end{figure}

The homogeneous steady state solution $\vh=(\eh,\sh,\ph)$ of \eqref{eq_SI:RDE_system} is given by the following expressions:
\begin{equation}\label{eq_SI:eh}
\eh = \gammas \frac{\sr - \sh}{\kcat F(\sh)},
\end{equation}
\begin{equation}
\ph = \frac{\kcat \eh F(\sh)}{\gammap},
\end{equation}
where $\eh$, $\sh$, and $\ph$ are the homogeneous concentrations of enzyme, substrate, and product respectively. In this work for simplicity we consider $\eh$ and $\ph$ as functions of $\sh$ but it is also possible to write $\sh$ and $\ph$ in terms of $\eh$, which is a quantity directly tunable in the experiments:
\begin{equation}\label{eq_SI:sh}
\sh = \frac{1}{2} \left[ \sr - \km - \frac{\kcat}{\gammas} \eh + \sqrt{ \left(\sr - \km - \frac{\kcat}{\gammas} \eh \right)^2 + 4 \sr \km} \right],
\end{equation}
\begin{equation}\label{eq_SI:ph}
\ph = \frac{\gammas}{2 \gammap} \left[ \sr + \km + \frac{\kcat}{\gammas} \eh - \sqrt{ \left(\sr - \km - \frac{\kcat}{\gammas} \eh \right)^2 + 4 \sr \km} \right],
\end{equation}
which for $\km \ll \sh < \sr$ become
\begin{equation}\label{eq_SI:sh-approx}
\sh = \sr - \frac{\kcat}{\gammas} \eh,
\end{equation}
and
\begin{equation}\label{eq_SI:eh-approx}
\ph = \frac{\kcat}{\gammap} \eh.
\end{equation}
In Fig.~\ref{fig:e-s-relation}, we plot $\eh$ versus $\sh$ for both the exact and the approximated relation. One can use Fig.~\ref{fig:e-s-relation} to read the enzyme concentration $\eh$ from the corresponding value of the substrate concentration $\sh$.

\subsection{Instability of the well-mixed system}

We first study the instability of the solution for the well-mixed system. In the well-mixed system, the spatial derivatives and any spatial dependence of the concentrations are neglected.
The system~\eqref{eq_SI:RDE_system} becomes:
\begin{equation}\label{eq_SI:RDE_system_WM}
\begin{aligned}
	\begin{cases}
	&\partial_t e(t) = 0 \\
	&\partial_t s(t) = -\kcat e(t) F(s) - \gammas (s - \sr) \\
	&\partial_t p(t) = \kcat e(t) F(s)- \gammap p \, .
	\end{cases}
\end{aligned}
\end{equation}
The fixed-point of the system~\eqref{eq_SI:RDE_system_WM} is still given by $\vh$. To perform a linear instability analysis, we perturb the system around its fixed point. Any perturbation around $\vh$ can be written as $e(t) = \eh + \delta e(t)$, $s(t) = \sh + \delta s(t)$, $p(t) = \ph + \delta p(t)$ or alternatively $\vec{\text{v}}(t) = \vh + \dv(t)$ where $\dv(t) = (\delta e(t), \delta s(t), \delta p(t))$. The amount of enzymes in the system is fixed and does not change over time because $\partial_t e(t)=0$, hence any perturbation $\delta e(t)$ just shifts the total enzyme amount.
The linearized  system of equation takes the form:
\begin{align}\label{eq_SI:well_mixed_linearized_RDE_system}
	\begin{cases}
	&\partial_t \delta e(t) = 0 \\
	&\partial_t \delta s(t) = - \kcat \eh F'(\sh) \delta s - \kcat F(\sh) \delta e(t) - \gammas \delta s(t)   \\
	&\partial_t \delta p(t) = \kcat \eh F'(\sh) \delta s + \kcat F(\sh) \delta e(t) - \gammap \delta p(t) \,,
	\end{cases}
\end{align}
where $F'(\sh)$ is the derivative $F(s)$ with respect to $s$ at $s=\sh$
\begin{equation}
F'(\sh) = \left.\frac{\partial F(s)}{\partial s}\right|_{s=\sh}.
\end{equation}
We rewrite the linear system of equations~\eqref{eq_SI:well_mixed_linearized_RDE_system} in matrix form
\begin{equation}\label{eq_SI:linear-dynamics}
\frac{\partial \dv(t)}{\partial t} = \tensor{J}_0 \cdot \dv(t),
\end{equation}
where
\begin{equation}\label{eq_SI:jacobian_well-mixed}
\tensor{J}_0 = \begin{pmatrix}
0 	&	0	&	0 \\
-\kcat F(\sh) 	&	-\kcat \eh F'(\sh) - \gammas	&	0 \\
\kcat F(\sh)	&	\kcat \eh F'(\sh)		&	- \gammap
\end{pmatrix}.
\end{equation}

The eigenvalues of $\tensor{J}_0$ determines the fate of the perturbations. The eigenvalues are $\sigma_1 = 0$, $\sigma_2 = -\kcat \eh F'(\sh) - \gammas$ and $\sigma_3 = -\gammap$ and all of them are non-positive. This indicates that the perturbations do not grow with time. $\sigma_1 = 0$ corresponds to the perturbation in the total enzyme amount. Once the total amount of enzymes is perturbed, its level does not change over time because $\partial_t e(t)=0$. The other eigenvalues are negative, i.e. any perturbations $\delta s(t)$, $\delta p(t)$ decay over time as long as $F'(\sh)>0$. Hence the homogeneous solution of a well-mixed system is stable against perturbations for any value of the parameters. 

\subsection{Instability of the reaction-diffusion system}

Similar to the well-mixed system, we again study the dynamics of small perturbations around the homogeneous steady state solution but the perturbations are now temporal and spatial dependent: $e(x,t) = \eh + \delta e(x,t)$, $s(x,t) = \sh + \delta s(x,t)$, $p(x,t) = \ph + \delta p(x,t)$ or alternatively $\vec{\text{v}}(x,t) = \vh + \dv(x,t)$ where $\dv(x,t) = (\delta e(x,t), \delta s(x,t), \delta p(x,t))$. We insert the perturbed concentrations in Eq.~\eqref{eq_SI:RDE_system} and we linearize the system by assuming that perturbations are small and find that:
\begin{align}\label{eq_SI:linearized_RDE_system}
	\begin{cases}
	&\partial_t \delta e(x,t) = \De(\sh) \partial_x^2 \delta e(x,t) + \left[ \eh \De'(\sh)  + \Dxd(\sh,\eh) \right] \partial_x^2 \delta s(x,t) \\
	&\partial_t \delta s(x,t) = \Ds \partial_x^2 \delta s(x,t) - \kcat \eh F'(\sh) \delta s - \kcat F(\sh) \delta e(x,t) - \gammas \delta s(x,t)   \\
	&\partial_t \delta p(x,t) = \Dp \partial_x^2 \delta p(x,t) + \kcat \eh F'(\sh) \delta s + \kcat F(\sh) \delta e(x,t) - \gammap \delta p(x,t) \,,
	\end{cases}
\end{align}
where $\De'(\sh)$, $F'(\sh)$ are the derivatives of $\De(s)$, $F(s)$ with respect to $s$ at $s=\sh$
\begin{equation}
\De'(\sh) = \left. \frac{\partial \De(s)}{\partial s}\right|_{s=\sh},
\end{equation}
\begin{equation}
F'(\sh) = \left.\frac{\partial F(s)}{\partial s}\right|_{s=\sh}.
\end{equation}
We write the linear system of equations~\eqref{eq_SI:linearized_RDE_system} in Fourier space to partially diagonalize the equations. The result is that the dynamics of each Fourier mode is independent from other modes and is governed by the following equation:
\begin{equation}
\frac{\partial \vec{\text{v}}(q,t)}{\partial t} = \tensor{J}(q) \cdot \dv(q,t),
\end{equation}
where
\begin{equation}
\dv(q,t) = \begin{pmatrix}
\delta e(q,t) \\
\delta s(q,t) \\
\delta p(q,t)
\end{pmatrix}
\end{equation}
and
\begin{equation}\label{eq_SI:jacobian}
\tensor{J}(q) = \begin{pmatrix}
-\De(\sh) q^2 	&	-[\Dxd(\sh,\eh) + \eh \De'(\sh)] q^2	&	0 \\
-\kcat F(\sh) 	&	-\kcat \eh F'(\sh) - \Ds q^2 - \gammas	&	0 \\
\kcat F(\sh)	&	\kcat \eh F'(\sh)		&	-\Dp q^2 - \gammap
\end{pmatrix}.
\end{equation}

It is not surprising that $\tensor{J}(0)$ is the same as the Jacobian of the well-mixed system $\tensor{J}_0$, \eqref{eq_SI:jacobian_well-mixed}. At $q=0$ we are neglecting all the effects due to diffusion and cross-diffusion, cf. Eq.~\eqref{eq_SI:jacobian_well-mixed}. Moreover perturbations with $q=0$ corresponds to homogeneous shifts in the concentrations, which are the same as considered for the well-mixed system.

The stability of the homogeneous solution can be determined by the eigenvalues of $\tensor{J}(q)$. If the real part of all the eigenvalues of $\tensor{J}(q)$ are negative, the homogeneous steady state solution is linearly stable; it is unstable otherwise. One of the eigenvalues of $\tensor{J}(q)$, $\sigma_3 = - \Dp q^2 - \gammap$, is always negative and corresponds to the relaxation rate of any perturbation that only perturbs the product concentration. From Eq.~\eqref{eq_SI:RDE_system} we can see how the product dynamics has no feedback on the substrate and enzyme equations. This implies that the instability is characterized by feedbacks in the $(e,s)$ subspace. The other two eigenvalues $\sigma_1$ and $\sigma_2$ are the solution of the following quadratic equation:
\begin{equation}
\begin{aligned}
&\sigma^2 + \sigma \left[ \kcat \eh F'(\sh) + \Ds q^2 + \De(\sh) q^2 + \gammas \right] \\
&+ \kcat \eh F'(\sh) \Df q^2  - \kcat F(\sh) \Dxd(\sh,\eh) q^2 + \gammas D_e (\sh) q^2 + \Ds \De(\sh) q^4 = 0\,.
\end{aligned}
\end{equation}
Using the definition of $\Dxd$, $\De$ and Eq.~\eqref{eq_SI:eh} we can write the following
\begin{equation}
\begin{aligned}
&\sh (\km + \sh) \sigma^2 + \sigma \left[ \gammas (\sr \km + \sh^2) + \sh q^2 (\km [\Ds + \Df] + \sh [\Ds + \Dc] ) \right] \\
&+ \gammas \left[ \km \sr \Df + \sh^2 \Dc + \sh (\sr - \sh) (\km \Cf + \sh \Cc) \right] q^2 + \sh \Ds [\km \Df + \sh \Dc] q^4 = 0.
\end{aligned}
\end{equation}
By rewriting the above expression as $(\sigma-\sigma_1)(\sigma-\sigma_2)=0$, we see that the summation of the eigenvalues is negative, therefore the smaller eigenvalue $\sigma_2$ is always negative. The larger eigenvalue $\sigma_1$ is positive if and only if the product of the two eigenvalues is negative. The condition for having negative product can be written as a condition for the wave vector ($q < q^*$), where
\begin{equation}\label{eq_SI:qc}
q^{*2} = - \frac{ \gammas \left[ \km \sr \Df + \sh^2 \Dc + \sh (\sr - \sh) (\km \Cf + \sh \Cc) \right] }{\Ds \sh [\km \Df + \sh \Dc]}.
\end{equation}
In order to have a positive $q^{*2}$, we should have that:
\begin{equation}\label{eq_SI:condition}
\Df + \left( \frac{\sh}{\km} \right) \left( \frac{\sh}{\sr} \right) \Dc + \frac{\sh}{\km} \left(1 - \frac{\sh}{\sr}\right) (\km \Cf + \sh \Cc) < 0.
\end{equation}
Equations \eqref{eq_SI:qc} and \eqref{eq_SI:condition} are the instability conditions for the system of equations~\eqref{eq_SI:RDE_system} (Eq.~(3) of the main text). In the regime $\sh \ll \sr$, with $\beta=\gammas \sr$, we get:
\begin{eqnarray}
& 0<q<q^* = \sqrt{- \frac{ \beta \left[ \km \Df + \sh (\km \Cf + \sh \Cc) \right] }{\Ds \sh [\km \Df + \sh \Dc]}}\,, \label{eq_SI:inst_cond_q} \\
& \Df +\sh \left( \Cf+ \frac{\sh}{\km} \Cc \right)<0 \,, \label{eq_SI:inst_cond_Df}
\end{eqnarray}
which are identical to the relations (4) and (5) of the main text.

By substituting $\Df = k_B T / 6 \pi \eta \Rf$, and $\Ccf = N_A k_B T \lambdacf^2 / \eta$ into relation \eqref{eq_SI:inst_cond_Df}, we obtain the analogous expression:
\begin{equation}\label{eq_SI:cond_lambdas}
\frac{1}{6 \pi N_A \Rf} \frac{\km}{\sh^2} + \frac{\km}{\sh} \lambdaf^2 + \lambdac^2 < 0.
\end{equation}
In case $\lambdac=\lambdaf$, by rearranging, we obtain the inequality (6) of the main text:
\begin{equation}\label{eq_SI:cond_constant_lambda}
\lambda^2 < -\frac{1}{6 \pi \Rf N_A}\frac{\km}{\sh(\sh+\km)}\,.
\end{equation}

To be able to see the patterns, the size of the system needs to be large enough to accommodate Fourier modes with $q < q^*$. Alternatively, the minimum system size to see patterns $L_{\min} = 2 \pi / q^*$:
\begin{equation}\label{eq_SI:L_min}
L_{\min} = 2\pi \sqrt{- \frac{\Ds \sh [\km \Df + \sh \Dc]}{\gammas \left[ \km \sr \Df + \sh^2 \Dc + \sh (\sr - \sh) (\km \Cf + \sh \Cc) \right]}}\,,
\end{equation}
where we considered the $q^*$ as given by Eq.~\ref{eq_SI:qc}.
For parameters as given by Table~\ref{tbl:fig2_si} (the same used for the simulation shown in Fig.~2A of the main text), $L_{\min}=59.2\mu m$.

\subsection{Instability type II}
\begin{figure}[!htb]
	\centering
	\includegraphics{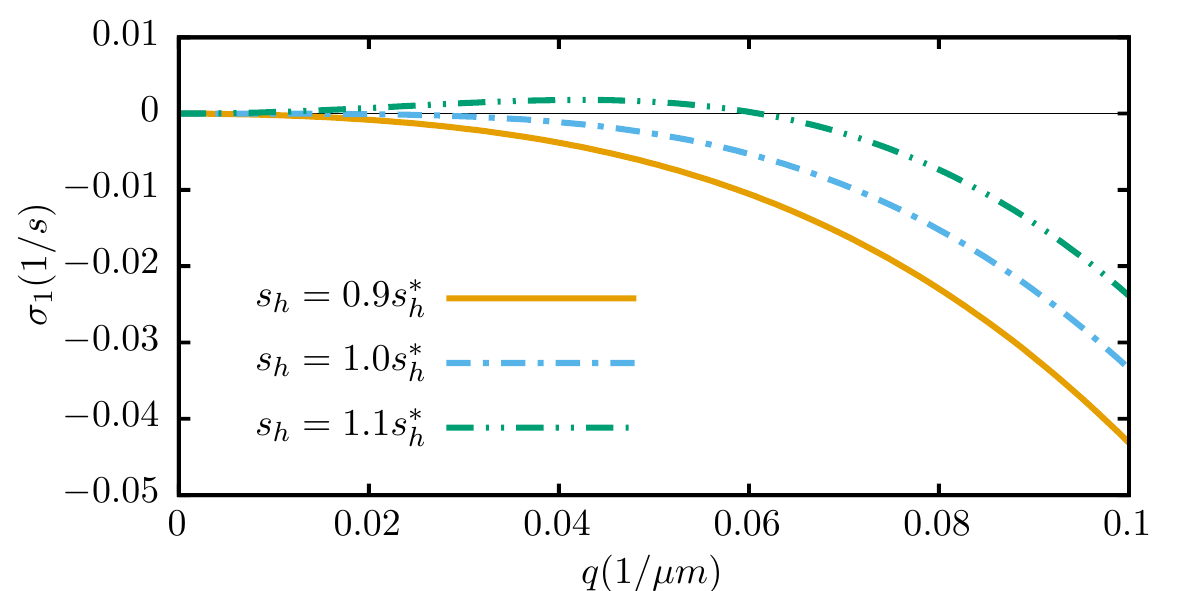}
	\caption{The largest eigenvalue of the Jacobian $\sigma_1$ versus the wave vector $q$. The horizontal black line corresponds to $\sigma_1 = 0$. By increasing the initial homogeneous concentration of substrate, $\sh$, from below to above critical concentration, $\sh^*$, the eigenvalues become positive.  The shape of the eigenvalue curves indicates that the instability is of type II. The parameters for the graph are $T=300K$, $\Df = 10 \mu m^2 / s$, $\Dc = 13 \mu m^2 / s$, $\Ds = 100 \mu m^2 / s$, $\km = 1 \mu M$, $\beta = 10^6 1/s$, $\lambdaf^2 = 1 \si{\angstrom}^2$, $\lambdac^2 = -1 \si{\angstrom}^2$}
	\label{fig:sigma}
\end{figure}
In this section we determine the type of instability for our system. As can be seen by looking at inequality \eqref{eq_SI:cond_lambdas}, we need repulsive nonspecific interactions to have an unstable system. Let us consider for example $\Cc<0$, then the inequality~\eqref{eq_SI:cond_lambdas} is fulfilled when $\sh > \sh^*$ where
\begin{equation}
\sh^* = \frac{-\km \lambdaf^2 - \sqrt{\km^2 \lambdaf^4 - \frac{2 \km}{3 \pi \Rf}}}{2 \lambdac^2}.
\end{equation}

Figure \ref{fig:sigma} shows the largest eigenvalue of the linearized equations. When $\sigma_1 > 0$, the system of equations is linearly unstable with respect to perturbations of the homogeneous solution. Fig.\ref{fig:sigma} indicates that the system gets unstable by increasing $\sh$ above its critical value, $\sh^*$, which is related indirectly to the initial amount of enzymes in the system. The form of curves in Fig.\ref{fig:sigma} correspond to a type II instability \cite{cross_pattern_2009}. A type II instability is typical of systems with conserved quantities. In our case the total amount of enzymes is conserved, this implies that $\sigma_1=0$ at $q=0$, otherwise we would have homogeneous change in $\eh$ over time, corresponding to changes in the total amount of enzymes.

\section{No short range interactions}
\label{SI_sec:No_short_range}
In this section we consider $\Dxd(s,e) = 0$. The Jacobian \eqref{eq_SI:jacobian} then takes the form:

\begin{equation}\label{eq_SI:jacobian_no_xD}
\tensor{J}(q) = \begin{pmatrix}
-\De(\sh) q^2 	&	-\eh \De'(\sh) q^2	&	0 \\
-\kcat F(\sh) 	&	-\kcat \eh F'(\sh) - \Ds q^2 - \gammas	&	0 \\
\kcat F(\sh)	&	\kcat \eh F'(\sh)		&	-\Dp q^2 - \gammap
\end{pmatrix}.
\end{equation}
$\sigma_3=-\Dp q^2 - \gammap$ is one of the eigenvalues and it is negative. The other two eigenvalues are obtained by solving the following equation:
\begin{equation}
\begin{aligned}
&\sigma^2 + \sigma \left( \gammas + \kcat \eh F'(\sh) + [ \De(\sh) + \Ds ] q^2 \right) \\
+& \left( \gammas \De(\sh) + \kcat \eh \left[ F'(\sh) \De(\sh) - \De'(\sh) F(\sh) \right] + \De(\sh) \Ds q^2 \right) q^2 = 0.
\end{aligned}
\end{equation}
By rewriting the above expression as $(\sigma-\sigma_1)(\sigma-\sigma_2)=0$, we see that the sum of the two eigenvalues is negative, as $F'(\sh) > 0$. The largest eigenvalue is positive if the product of the two eigenvalues is also negative. Thus the term with zeroth order of $\sigma$ should be negative. Together with the identity Eq.~\eqref{eq_SI:eh}, i.e. $\kcat \eh = \gammas (\sr - \sh) / F(\sh)$, we find the following condition:
\begin{equation}\label{eq_SI:no-phoretic-condition}
\left[ \frac{\De'(\sh)}{\De(\sh)} - \frac{F'(\sh)}{F(\sh)} \right] > \frac{1}{\sr - \sh} + \frac{\Ds q^2}{\gammas (\sr - \sh)}.
\end{equation}

In the regime $\sh \ll \sr$, with $\beta=\gammas \sr$, we get:
\begin{equation}
\left[ \frac{\De'(\sh)}{\De(\sh)} - \frac{F'(\sh)}{F(\sh)} \right] > \frac{\Ds q^2}{\beta},
\end{equation}
which is the same as the inequality~(7) of the main text. 

By using the definition of $\De(s)$ used in \cite{agudo-canalejo_phoresis_2018} $\De = \Df + (\Dc - \Df) F(s)$, we find that the inequality \eqref{eq_SI:no-phoretic-condition} takes the form:
\begin{equation}\label{eq_SI:contraddiction}
-\frac{\Df F'(\sh)}{\De F(\sh)} > \frac{1}{\sr - \sh} + \frac{\Ds q^2}{\gammas (\sr - \sh)}.
\end{equation}
Because $\sr > \sh$ and $F'(\sh) > 0$, the left hand side of the inequality above is negative and cannot be larger than the always positive right hand side. Therefore enhanced diffusion alone cannot drive instabilities. The cross-diffusion generated by repulsive interactions is key in this respect. This also holds in the simpler case of $\sh \ll \sr$, with $\beta=\gammas \sr$.

Similarly it is possible to show that we cannot have instabilities for other enhanced diffusion definitions \cite{jee_enzyme_2018,mohajerani_theory_2018}.

\section{Parameters of the simulations}
\label{SI_sec:params}
The numerical simulations of the system of equations~\ref{eq_SI:RDE_system} have been carried out by using the COMSOL Multiphysics v5.3.

We used the parameters listed in Table~\ref{tbl:fig2_si} for the simulations shown in Fig.~2 of the main text. We used $\lambdaf^2=\lambdac^2=-1\,\si{\angstrom^2}$ for the unstable homogeneous solution (upper panel) and $\lambdaf^2=\lambdac^2=1\,\si{\angstrom^2}$ for the stable homogeneous solution (lower panel). 
\begin{table}
	\caption{Parameters used for the COMSOL simulations shown in Fig.~2 of the main text}
	\label{tbl:fig2_si}
	\begin{tabular}{|ll||ll|}
		\hline
		Parameter  & Value &  & \\
		\hline
		$\Df$					& $10$ $\mu m ^2/s$			 	 & $\Dc$						& $13 $ $\mu m^2/s$ 		\\
		$\Ds$					& $100$ $\mu m^2/s$				 & $\Dp$						& $100$ $\mu m^2/s$ 		\\
		$\kcat$					& $10^4$ $1/s$					 & $\km$						& $1$ $\mu M$       		\\
		$\sh$					& $10^4$ $\mu M$				 & $\sr$						& $10^6$ $\mu M$	  		\\
		$\lambdaf^2$ (stable)	& $1$ $\si{\angstrom}^2$		 & $\lambdac^2$ (stable)		& $1$ $\si{\angstrom}^2$ 	\\
		$\lambdaf^2$ (unstable)	& $-1$ $\si{\angstrom}^2$		 & $\lambdac^2$ (unstable) 		& $-1$ $\si{\angstrom}^2$	\\
		$\gammas$				& $1$ $1/s$						 & $\gammap$					& $10$ $1/s$				\\
	    $\eta$				    & $8.9\cdot 10^{-4}$ $Pa\cdot s$ & $T$							& $300$ $K$					\\			
	    $L$						& $100$ $\mu m$ 				 &								&							\\
		\hline
	\end{tabular}
\end{table}
The initial homogeneous concentrations $\eh$, $\sh$, $\ph$ were perturbed by adding white Gaussian noise with variance $\eh/100$, $\sh/100$, $\ph/100$ respectively. We considered periodic boundary conditions and we used the ``Time dependent'' solver of COMSOL with a relative tolerance of $10^{-6}$ and an element size of $0.01\,\mu m$, i.e. $10^4$ lattice points. 

For the results shown in Fig~3 we again used the ``Time dependent'' solver of COMSOL. We considered a grid of $20\times 20$ values for $\lambdac^2=\lambdaf^2=\lambda^2$ in the interval $[-10,3]\si{\angstrom}^2$ (evenly spaced), and $\sh$ in $[10^1,10^5]\mu M$ (evenly spaced on a log-scale). For the COMSOL specific parameters, we used a relative tolerance of $10^{-6}$ and a maximum element size of $0.02\,\mu m$, except when two peaks were observed as a final result of the simulation. In these cases, we repeated the simulations with a finer grid (element size of $0.01\,\mu m$) and again we observed a single peak for the concentrations as the final result of the simulations. All the other parameters were the same as of Table~\ref{tbl:fig2_si} and the boudary conditions were periodic.

\end{document}